\documentclass[twocolumn,aps,prl,showpacs]{revtex4}

\usepackage{subfiles}

\usepackage{graphicx}

\usepackage{amsmath}

\begin{document}

\title{Nonlinear conductance of long quantum wires at a conductance plateau transition: 
Where does the voltage drop?}

\author{T.~Micklitz$^1$, A.~Levchenko$^2$, and A.~Rosch$^3$}

\affiliation{$^1$Dahlem Center for Complex Quantum Systems and Institut 
f\"ur Theoretische Physik, Freie Universit\"at Berlin, 14195 Berlin, Germany\\
$^2$Department of Physics and Astronomy, Michigan State University, East Lansing, MI 48824, USA\\
$^3$Institut f\"ur Theoretische Physik, Universit\"at zu K\"oln, Z\"ulpicher Str. 77, 50937 Cologne, Germany}

\date{February 19, 2012}


\begin{abstract}

We calculate the linear and nonlinear conductance of spinless 
fermions in clean, long quantum wires where short-ranged interactions 
lead locally to equilibration. Close to the quantum phase transition 
where the conductance jumps from zero to one conductance quantum, 
the conductance obtains an universal form governed by the ratios of 
temperature, bias voltage and gate voltage. Asymptotic analytic results 
are compared to solutions of a Boltzmann equation which includes the 
effects of three-particle scattering. Surprisingly, we find that for long wires 
the voltage predominantly drops close to one end of the quantum wire due 
to a thermoelectric effect.

\end{abstract}

\maketitle

\textit{Introduction.---} A clean quantum wire with adiabatic contacts is 
characterized by a quantized conductance, $G=n G_0$ with $G_0=e^2/h$. 
The integer $n$ describes the number of conduction channels (including spin). 
The conductance quantization is closely related to charge quantization and 
survives (for sufficiently low temperatures $T$) even in the presence of 
interactions~\cite{maslov,ponomarenko,safi} as long as momentum relaxation 
by Umklapp scattering can be neglected~\cite{roschUmklapp,giamarchi}.

The transition from one conductance plateau to the next is an example of a 
quantum phase transition without order parameter, where only a topological 
property, the number of conducting channels, changes. While the thermodynamics 
of this quantum critical point (QCP) is quite well understood~\cite{sachdevBook,balents,QPTgate,TSBQW}, 
a theory of the quantum critical conductance is much more challenging: 
How does the conductance change from one conductance plateau to the 
next at low but finite $T$? We will answer this question both in the linear 
and non-linear regime for the most simple situation, i.e., the transition from 
$n=0$ to $n=1$ for spinless fermions with finite-ranged interactions. 
Here the QCP describes the transition from zero to a finite fermion density. 
As interactions are irrelevant at this QCP, thermodynamic properties are 
well described by non-interaction fermions~\cite{CHNCuBr,giamarchi} but 
transport in long wires is still governed by collisions. Relaxation by collissions 
and non-equilibrium dynamics in one-dimensional (1D) systems have recently 
moved into the focus of theoretical~\cite{mirlin,gutman1,gutman2,takei,torsten,IG,Khodas,lunde} 
and experimental~\cite{altimiras, chen, barak} research.

A general question is where the voltage drops when a finite current is driven through a 
clean 1D quantum wire by applying a bias voltage $V$. For diffusive (multichannel) 
quantum wires one expects a linear drop of the voltage (i.e. of the electrochemical potential) 
across the wire while for non-interacting, ballistic quantum wires the voltage drop occurs 
only close to the two contacts~\cite{imry}. In clean {\em interacting} quantum wires with 
low fermion density (and therefore negligible Umklapp scattering) the dc conductivity 
is infinite for an infinitely long wire due to momentum conservation, $\sigma(T)=\infty$. 
The vanishing resistivity strongly suggests that there is again no voltage drop inside the wire.

A recent series of papers~\cite{{feq,peq,eqLL,GLL,eqWC}}, which studied the role of 
equilibration in long but finite quantum wires of length $L$, found that in the linear 
response regime, $V \to 0$, there is a linear drop of voltage~\cite{peq} along the wire. 
We resolve this apparent contradiction to $\sigma(T)=\infty$ by noting that the limits 
$V \to 0$ and $L \to \infty$ do not commute. The drop of voltage  is governed by a new 
length scale $\ell_V$ which diverges for $V \to 0$. For $L \ll \ell_V$ a linear drop of voltage 
occurs. In the opposite limit, $L \gg \ell_V$, however, the voltage drops only within a 
distance of $\ell_V$ of the contacts. Surprisingly, the voltage drop is not symmetrical and 
occurs predominantly only at one of the two contacts!

Previous work on equilibrated quantum wires~\cite{feq,peq,eqLL,GLL,eqWC} focused 
on the limit $T \ll \epsilon_F$, where $\epsilon_F$ is the Fermi energy. As scattering 
processes equilibrating left-moving and right-moving fermions involve the bottom of 
the band~\cite{lunde,AZT,peq,eqLL,eqWC}, they are exponentially suppressed and 
therefore the corresponding equilibration length is exponentially large, 
$\ell_{\rm eq}\sim e^{\Delta/T}$, where $\Delta \approx \epsilon_F$ for weak interactions 
(thermal equilibration relevant for heat conductance occurs on shorter length scales~\cite{K}). 
For  $L \gg \ell_{\rm eq}$ it was found that the quantized (linear) conductance obtains 
corrections of order $(T/\Delta)^2$. Large effects can therefore  be expected close to the 
conductance plateau transition, where $\Delta \sim T$, as studied in this paper.

{\it Model.---} We consider 1D spin-polarized electrons with quadratic dispersion 
$\epsilon_p={p^2\over 2m}$, interacting via a short range potential. Close to the QCP, 
where filling of the first subband becomes small,  interactions are strongly irrelevant in 
the renormalization group sense~\cite{sachdevBook,supplement} and a single electron 
description becomes approximately valid. To study equilibration and its effect on transport 
we may thus use the Boltzmann equation
\begin{align}
\label{be}
v_p \partial_x f_{x,p} = - I_{x,p}^{\rm col}[f],
\end{align} where $f_{x,p}$ is the quasiclassical distribution function, $v_p=p/m$ the 
velocity, and  the collision integral $ I^{\rm col}$ describes collisions. The contacts of 
the quantum wire to the leads at $x=\pm \frac{L}{2}$ induce boundary conditions for 
electrons moving {\em into} the quantum wire
\begin{align}\label{boundary}
f_{x=-\frac{L}{2},p>0}= \frac{1}{ e^{ {\xi_p^l/ T}}+1 }, \quad
f_{x=\frac{L}{2},p<0}= \frac{1}{ e^{ {\xi_p^r/ T}}+1 },
\end{align} where $ \xi_p^{l/r}=\epsilon_p-\mu \mp eV/2$ with $\mu=0$ at the QCP. 
Here we assume adiabatic and ballistic contacts, i.e., contacts which are smooth 
compared to the electronic wavelength but short in comparison to the scattering length.

In 1D systems energy and momentum conservation severely restrict the phase 
space available for scattering: in a two-particle process, two particles of equal 
mass can only exchange their momenta~\cite{sachdevBook,lunde} which 
leaves $f_p$ unchanged. One therefore has to study the effects of three 
particle collisions~\cite{lunde,AZT} described by
\begin{align}
\label{ci}
I_{x,p_1}^{\rm col}[f]
=
\sum_{p_2p_3 \atop p'_1p'_2p'_3}
&
W_{123}^{1'2'3'} \big[
f_1f_2f_3(1-f_{1'})(1-f_{2'})(1-f_{3'}) \nonumber \\
- &
f_{1'}f_{2'}f_{3'}(1-f_1)(1-f_2)(1-f_3)
\big]
\end{align} where the scattering rate $W_{123}^{1'2'3'}$  arises to fourth order 
in the bare two-particle interactions \cite{fnW}. For low energies and spinless 
fermions, Pauli principle ensures that it takes the universal form
\begin{align}
\label{3pex}
& W_{123}^{1'2'3'}
=W
\big((p_1 - p_2) (p_1 - p_3) (p_2 - p_3) \nonumber \\
& \quad (p'_1 - p'_2 ) (p'_1 - p'_3) (p'_2 - p'_3 )\big)^2
\delta_{P_i,P_f}\delta(E_i-E_f)
\end{align}
where $P_{i(f)}=p_1+p_2+p_3$ and $E_{i(f)}=\epsilon_1+\epsilon_2+\epsilon_3$ 
are the total momentum and energy of the three scattering particles before (after) 
the collision, respectively. 
A simple dimensional analysis allows to identify a characteristic length scale of 
equilibration at the QCP ($\mu =0$) by setting typical momenta to $\sqrt{2 m T}$
\begin{align}
 \frac{1}{\ell_{\rm eq}}
=\frac{2W m^2L^4 }{(2\pi\hbar)^4}(2 m T)^{13/2} \label{leq}
\end{align} Measuring all length scales in units of  $\ell_{\rm eq}$ and all momenta 
in units of $\sqrt{2 m T}$  allows to scale out the parameters $W$, $m$ and $T$ 
and the only remaining parameters are $L/\ell_{\rm eq}$, $eV/T$ and $\mu/T$. 
We have checked both numerically and analytically that close to the QCP 
Hartree-Fock potentials (not included in Eq.~(\ref{be})) can be neglected.

For a numerical solution of the Boltzmann equation~(\ref{be}) it is important 
to avoid discretization errors leading to a violation of conservation laws. 
We therefore use a conservative splitting method following Ref.~\cite{Aristov}, 
see supplement \cite{supplement}, to solve the time-dependent Boltzmann 
equation until a steady state has been reached. For the linear response 
calculation we use a linearized collision integral.
\begin{figure}[bb]
\begin{center}
a) \includegraphics[width=72mm]{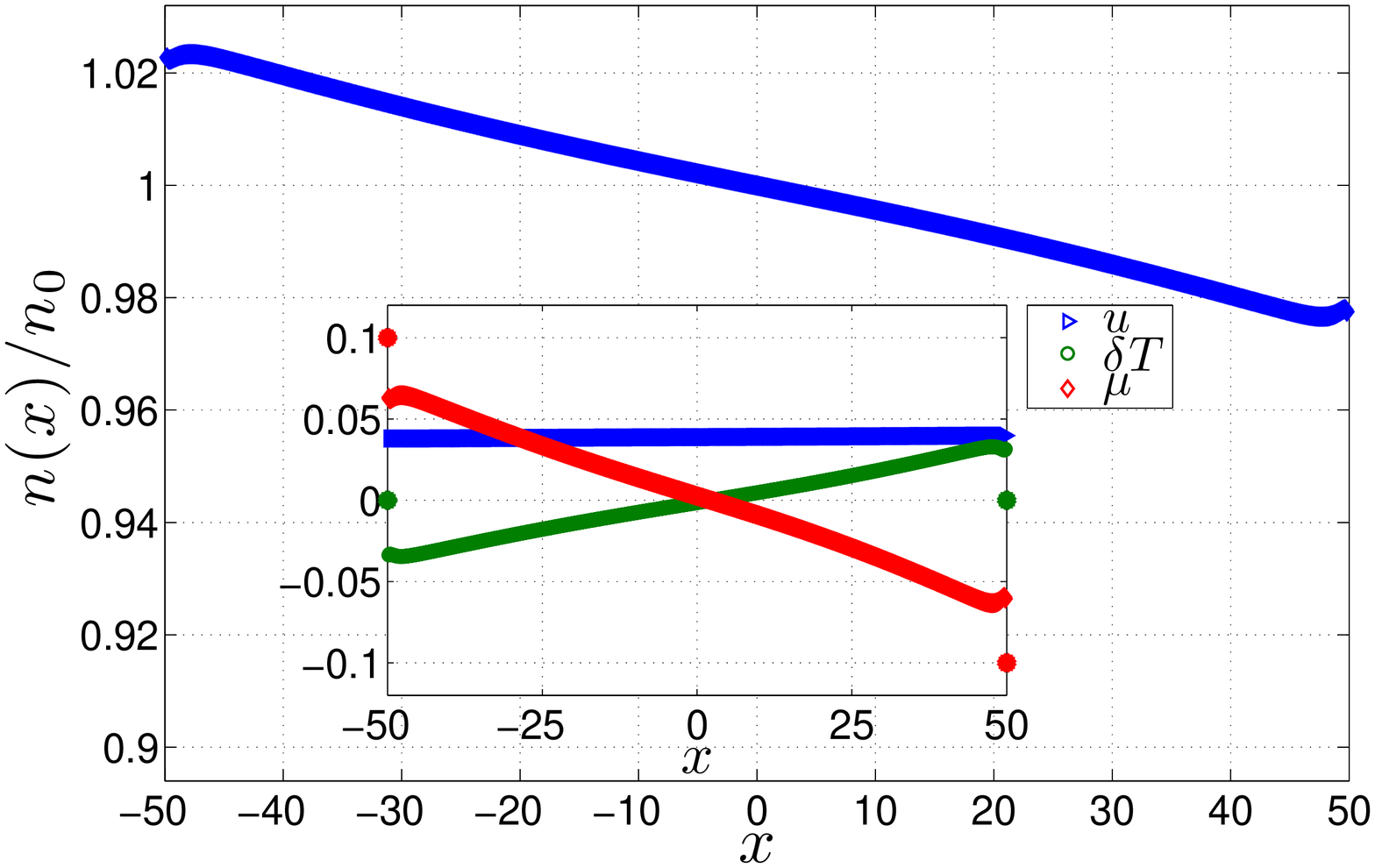} \\
b) \includegraphics[width=72mm]{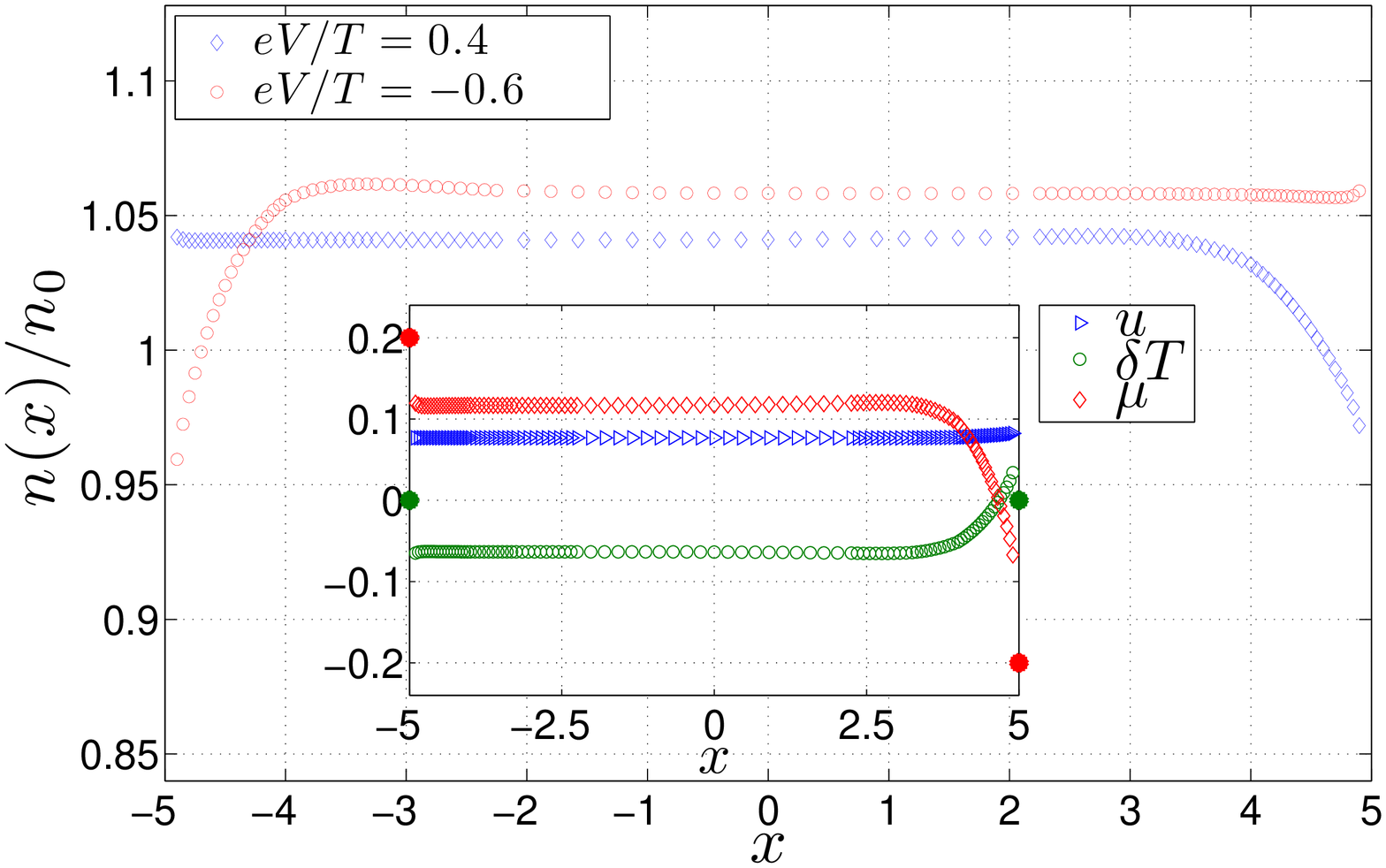}
\end{center}
\vspace{-15pt}
\caption{(Color online) Electron density, $n(x)=n_0(1+\delta n(x))$, at the 
QCP ($\mu=0$) calculated from a solution of the Boltzmann equation~(\ref{be}), 
using (a) a linearized collision integral for $L=100 \ell_{\rm eq}$ and $V \to 0$ 
and (b) the non-linearized collision integral for  $L=10 \ell_{\rm eq}$ and larger 
voltages, $eV/T=0.4$ and $eV/T=-0.6$ (inset: $eV/T=0.4$). The inset shows 
$\mu(x)$, $\delta T(x)=T(x)-T$ and $u(x)$ (in units of $T$ and $\sqrt{2T/m}$) obtained from 
fitting the local charge, energy and momentum densities to  Eq.~\eqref{feq}. 
While in the linear response regime, $L \ll |\ell_V|$, there is a linear voltage 
drop across the wire, the voltage (and the density) drops predominantly close 
to one of the two contacts for $L \gg |\ell_V|$. Note that due to a finite drift $u$ 
the chemical potentials close to the contacts do not match the chemical 
potential $\mu \pm eV/2$ in the leads (shown as separate dots at $x=\pm L/2$ 
in the insets).} \label{fig12}
\end{figure}

{\it Conservation laws.---} Three conservation laws govern transport in long 
quantum wires: charge, energy and momentum conservation. The corresponding 
currents are the charge current, $j_c= e\sum_p v_p f_p$, the energy current, 
$j_E \approx \sum_p \epsilon_p v_p f_p$, and the momentum current, 
$j_p \approx \sum_p p v_p f_p$. The latter can be identified with pressure. 
For sufficiently long quantum wires and far away from the contacts the system 
will reach locally equilibrium with the distribution function
\begin{align}
\label{feq} f^{\rm eq}_{p}(u,\mu,T)= \big[1+ e^{ {(p-mu)^2\over 2mT} -{\mu\over T} } \big]^{-1}
\end{align}
parametrized by three space-dependent Lagrange parameters $\mu(x)$, $T(x)$ 
and the velocity $u(x)$ reflecting the three conservation laws. For the distribution  
function (\ref{feq}) one can calculate the corresponding equilibrium currents 
$j_c^{\rm eq}$, $j_E^{\rm eq}$ and $j_p^{\rm eq}$ as function of $\mu$, $T$ and $u$.

{\it Voltage drop.---} Fig.~\ref{fig12} shows the density profile and 
the local chemical potential (insets) of long quantum wires, $L\gg
\ell_{\rm eq}$, obtained from our Boltzmann simulations (the 
supplement \cite{supplement} discusses how $\mu(x)$ and $T(x)$ can be 
measured by tunneling contacts). For small $V$ 
(Fig.~\ref{fig12}a) there is both a linear drop of the chemical 
potential along the wire and a finite jump directly at the two 
contacts (the separate points at $x=\pm L/2$ show $\mu$ in the leads).
In an experiment, this jump will occur on the length scale describing the 
crossover from the 1D lead to the higher-dimensional contacts.
This jump is also present for larger $V$ 
(Fig.~\ref{fig12}b), where, however, the linear voltage drop is 
absent. Surprisingly, there is instead a large {\em asymmetric} 
voltage drop which occurs only close to one of the two contacts. This 
behavior occurs for sufficiently long wires not only directly at the 
QCP but also away from it. Interestingly, one observes cooling 
  instead of heating close to the left contact. 
  Though unexpected at first glance, it can be related~\cite{peq} to 
  the finite boost $u$ in \eqref{feq}, which is partially compensated 
  by a reduced temperature to match the boundary condition for 
  right-movers.

The qualitative difference between small and larger voltage can be understood from 
a simple argument based on matching currents. The steady state for $V>0$ is 
characterized by the three currents $j_c$, $j_E$ and $j_p$. From the three 
equations $j_\alpha=j_\alpha^{\rm eq}$, $\alpha=c,E,p$, one can, for sufficiently 
long wires, determine the three parameters $\mu$, $T$ and $u$ which will be 
constant along the wire as $j_\alpha=const.$. Therefore, for a sufficiently long 
wire and finite $V$, a voltage drop can occur only close to the contacts. In the 
linear response regime, i.e. for small $V$, the situation is, however, different. 
By setting only one of the three parameters, $u$, to zero, two currents, 
$j_c^{\rm eq}$ and $j_E^{\rm eq}$, vanish in equilibrium. As both $j_c^{\rm eq}$ 
and $j_E^{\rm eq}$ are linear in $u$, their ratio $j_c^{\rm eq}/j_E^{\rm eq}$ 
is -- in the limit of small $V$ -- fixed by the average $\mu$ and $T$. 
This is used below when calculating the linear-response conductance analytically.

To develop an approximate analytical theory valid in both regimes, we consider 
small, but finite voltages $V$ and parametrize $f_{x,p}$ by
\begin{align}
\label{feqdf}
f_{x,p}=f^{\rm eq}_{x,p} +\delta f_{x,p},
\end{align}
where $\delta f_{x,p}$ accounts for deviations from local equilibrium 
$f^{\rm eq}_{x,p}=f^{\rm eq}_{p}(u(x),\mu(x),T(x))$. 
Here it is convenient to determine $\mu(x)$, $T(x)$ and $u(x)$ from the two equations 
$j_c=j_c^{\rm eq}$ and $j_p=j_p^{\rm eq}$ while the third parameter is fixed by fitting 
the local density $n(x)=n^{\rm eq}(x)$.

By linearizing the Boltzmann equation (\ref{be}) in $\delta f$, one obtains that 
$\delta f$ is proportional to $\partial n/\partial x$. For the energy current, one 
therefore obtains
\begin{eqnarray}\label{hydro0}
j_E=j_E^{\rm eq}+\tilde D \frac{\partial n }{\partial x}=const.
\end{eqnarray}
where $\tilde D$ is 
the thermoelectric diffusion constant describing how density gradients generate energy currents. 
Using Kubo's formula and Einstein relations, $\tilde D$ can be
calculated from the product of a correlation
function of the heat- and particle current and the compressibility.

For small voltages $\tilde D$ 
is approximately constant across the wire. Using Galilei invariance which implies 
$u=j_c^{\rm eq}/en^{\rm eq}=j_c/en$, we obtain 
$j_E^{\rm eq}=\frac{3  j_p j_c }{2en}- \frac{m{j_c}^3}{e^3n^2} $. 
For small $V$ the last term can be neglected and one can linearize the density 
$n=n_0+\delta n$ to obtain
\begin{eqnarray}\label{hydro}
-\frac{3  j_p j_c }{2 e n_0^2} \delta n
+\tilde D \frac{\partial \delta n }{\partial x}\approx j_E- \frac{3  j_p j_c }{2 e n_0}=const.
\end{eqnarray}
This equation  introduces a new length scale
\begin{eqnarray}
\ell_V=\frac{2 \tilde D e n_0^2}{3  j_p j_c }
\end{eqnarray}
which diverges for $V\to 0$ as $j_c$ vanishes in this limit while $\tilde D$, $n_0$ and 
$j_p$ remain finite. For $|eV| \ll T$ and $\mu=0$, i.e., at the QCP,  a simple dimensional 
analysis gives  $\ell_V \sim \ell_{\rm eq} \frac{T}{eV}$.

For $L \ll |\ell_V|$, one obtains from Eq.~(\ref{hydro}) 
$\frac{\partial \delta n }{\partial x}=const.$ and therefore a linear drop in density and 
local chemical potential as in our numerical results, Fig.~\ref{fig12}a. In the other limit,  
$L \gg |\ell_V|$, $\delta n$ obtains an exponential $x$ dependence 
\begin{align}
\label{nfe}
n(x) = n_L + (n_R-n_L) \exp\!\left[ \frac{x-L/2}{\ell_V}\right]
\end{align} with $n_{L/R} \approx n(\mp L/2)$. 
The direction of the current 
determines whether the drop of density and voltage occur at the 
right ($\ell_V>0$) or left ($\ell_V<0$) lead, see  Fig.~\ref{fig12}b. 
This shows that the strongly asymmetric drop of voltage arises from a thermoelectric  
effect captured by the simple hydrodynamic equation (\ref{hydro}).

{\it Linear response regime.---} Interestingly, it is possible to calculate in the linear 
response regime the quantum critical conductance  for long quantum wires 
($\ell_{\rm eq} \ll L \ll |\ell_V|$) analytically. We use the approach developed 
in Ref.~\cite{feq,peq} (where only $T \ll \mu$ was considered) and keep track 
of the change of the charge and energy current carried by {\em right-moving} 
electrons with $p>0$, $j^R_c(x)=e\sum_{p>0} v_p f_{x,p}$ and 
$j^R_E(x)=\sum_{p>0}  v_p \epsilon_p f_{x,p}$, respectively. 
We use that far away from the contacts the distribution function obtains local 
equilibrium, $f_{x,p}\approx f^{\rm eq}_{x,p}$, described by Eq.~(\ref{feq}) with 
$T(x)=T+\delta T(x)$,  $\mu(x)=\mu+\delta \mu(x)$ and $u(x)$. This 
allows
to calculate directly  $j_c$, $j_c^R$, $j_E$, $j_E^R$ and $j_p$ in 
terms of three unkown functions, 
$u(x)$, $\delta \mu(x)$ and $\delta T(x)$. 
Current conservation implies $u(x)=const$. Furthermore, the ratio $r_1=j_E/j_c$ 
is to linear order just a simple function of 
$\mu$ and $T$ independent of $V$ and $u$. 
 The condition of constant momentum current  
fixes another ratio, $r_2=\partial_x\delta T/\partial_x\delta 
\mu$. This result is used to eliminate all unkowns from the ratio  
$r_3=\partial_x j^R_E/\partial_x j^R_c$.  
To leading order, $r_1, r_2$ and $r_3$ are space independent functions of 
$\mu$ and $T$ (calculated in the supplementary material~\cite{supplement}).  
Finally, one identifies~\cite{feq} the difference in the charge (energy) current 
of the interacting- and non-interaction system as the total change in the 
right-moving charge  (energy) current along the wire, 
$j_c=j_c^0+\int \partial_x j^R_c dx$ ($ j_E=j_E^0+\int \partial_x j^R_E dx$),  
respectively. If we now assume (as we checked numerically), that these 
integrals are dominated by their bulk contribution, we obtain the equation 
\begin{eqnarray}
r_3=\frac{r_1 j_c-j_E^0}{j_c-j_c^0}
\end{eqnarray}
from which one can calculate directly $j_c$. Combining all 
results~\cite{supplement}, we find for the linear-response 
conductance up to corrections of ${\cal O}(\ell_{\rm eq}/L)$, i.e. for 
$\ell_{\rm eq} \ll L \ll |\ell_V|$ 
\begin{align}
\label{gz}
G(z) = {e^2\over h} {
\alpha_0(z)\alpha_2(z)-\alpha_1^2(z)
\over
\alpha_2(z) + \alpha_0(z) \kappa^2(z)  - 2\alpha_1(z)\kappa(z)
},
\end{align}
where $z=\mu/T$, $\langle ... \rangle_{z}= - \int_{-{z}}^{\infty}  
d\xi (...) {d f^0_\xi  \over d\xi}$, with $f^0_{\xi} ={1\over 1 + e^{\xi} }$  
and $\alpha_k =\langle \xi^k \rangle_{z}$, 
$\kappa = { \langle \xi \sqrt{z + \xi } \rangle_{z} \over \langle \sqrt{z +  \xi }\rangle_{z}}$. 
At the QCP, i.e., for $\mu=0$, this gives 
\begin{align}
\label{gqcp}
\frac{G_{\rm QCP}}{e^2/h}=  { {\pi^2\over6}-2\ln^2 2 \over
{\pi^2\over 3}+{9\over 8}{\zeta^2(3/2) \over \zeta^2(1/2)}
+ {6\over \sqrt{2}}{\zeta(3/2)\over \zeta(1/2)} \ln 2 }\approx 0.420
\end{align} with $\zeta(x)$ the Riemann zeta function.  
$G_{\rm QCP}$ is about $16\%$ below the non-interacting result $e^2/2 h$.

\begin{figure}[bb]
\begin{center}
\includegraphics[width=8cm]{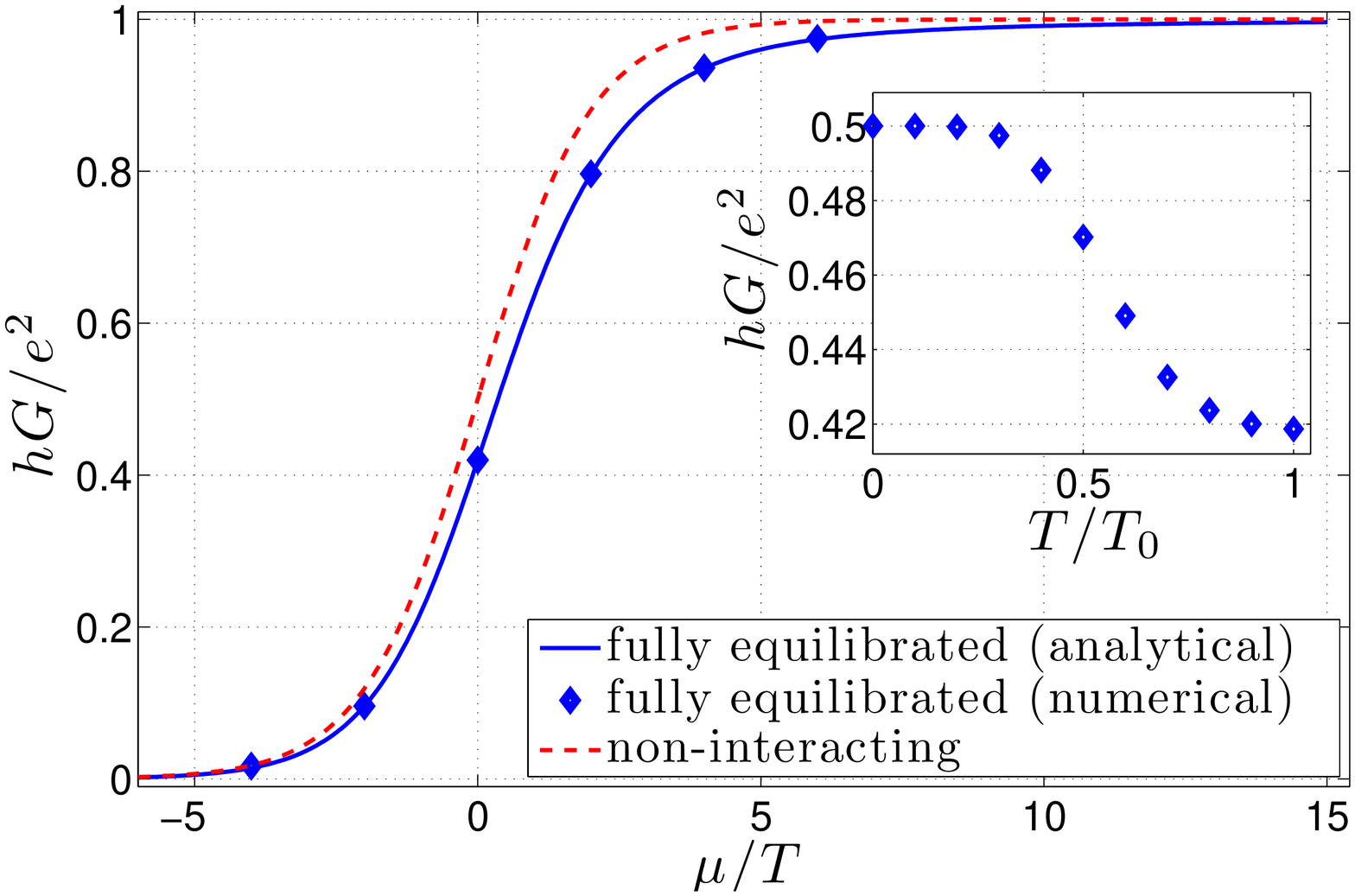}
\end{center}
\vspace{-15pt}
\caption{(Color online) Conductance of fully equilibrated (solid line) and 
non-interacting electrons (dashed line) in the linear response regime 
($\ell_{\rm eq} \ll L \ll |\ell_V|$). Numerical results (symbols) agree with 
Eq.~(\ref{gz}).  Inset: Upon lowering $T$, $\ell_{\rm eq}$ grows rapidly 
and a crossover from the equilibrated to the  non-interacting  conductance 
is observed for $\ell_{\rm eq} \sim L$ ($L=10\ell_{\rm eq}, \mu=0$).}
\label{fig3}
\end{figure} 
Fig.~\ref{fig3} displays the linear response conductance as functions of $\mu/T$ 
for non-interacting (see below) and fully equilibrated electrons which have a 
clearly different shape. Our analytical formula (\ref{gz}) fits very well the 
numerical result (symbols).

{\it Shorter wires.---} Upon lowering $T$, $\ell_{\rm eq}$ rapidly increases, 
see Eq.~(\ref{leq}). For  quantum wires, where $L/\ell_{\rm eq} \ll 1$, one can 
neglect the effects of equilibrating interactions. Half of the voltage drops at the 
left and right contact, respectively, and there is no voltage drop inside the wire 
as $f_{x,p}=f_p$ is independent of $x$. For $j_c$ one obtains the well-known 
non-interacting result 
$j_c^0= \frac{eT}{h} \ln\left[\frac{1+e^{(\mu+eV/2)/T}}{1+e^{(\mu-eV/2)/T}}\right]$. 
The conductance plateau transition in linear response is therefore described by 
$G(\mu/T)=G_0/( 1+ e^{-\mu/T})$ while at $\mu=0$ the current is for arbitrary 
$eV/T$ given by $j_0=G_0 V/2$. In the inset of Fig.~\ref{fig3} we have calculated 
numerically  the crossover from the interacting quantum critical conductance 
(\ref{gqcp}) to the non-interacting one, which occurs  upon lowering $T$ when 
$\ell_{\rm eq} \sim L$.
\begin{figure}[tb]
\begin{center}
\includegraphics[width=82mm]{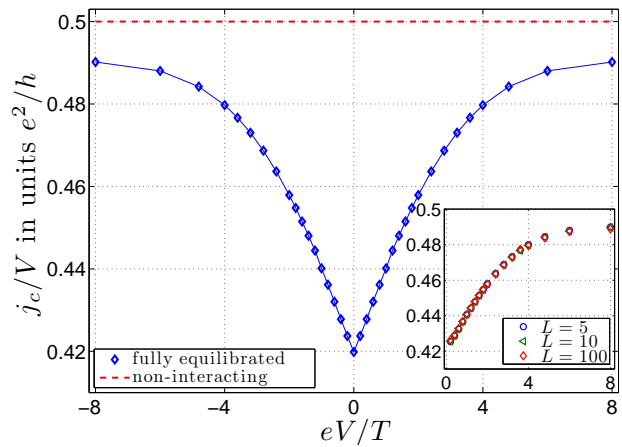}
\end{center}
\vspace{-15pt}
\caption{(Color online) Non-linear conductance, $j_c/V$ obtained for 
$L=7.5\ell_{\rm eq}$ from a numerical solution of the Boltzmann equation. 
For $eV/T\to 0$ a linearized collision integral was used. Inset: Within our 
numerical precision, there is no finite size dependence of the non-linear 
conductance for $L \gg \ell_{\rm eq}$ ($L$ measured in units $\ell_{\rm eq}$ ). }
\label{fig4}
\end{figure}

{\em Nonlinear response.---} Fig.~\ref{fig4} show the nonlinear conductance $j_c/V$ 
at the QCP, i.e. for $\mu=0$. It interpolates between the linear-response 
value (\ref{gqcp}) and the non-interacting result obtained for $|eV|/T \to \infty$. 
For $|eV| \gg T$ and $\mu=0$ all states with $p>0$ and $\epsilon_p<|eV|/2$ are 
occupied. As this is also an equilibrium distribution function with $ u=\sqrt{|eV|/4 m}$ 
and $\mu= m u^2/2$, collisions have no effects in this limit.

For small $V$ the nonlinear conductance appears to be non-analytic,
\begin{align}
j_c(V)=  G_{\rm QCP} V + \gamma |V| V + ...
\end{align}
which can be traced back to the asymmetric voltage drop for $L \gg |\ell_V|$. As 
$n_R-n_L$ in Eq.~(\ref{nfe}) varies linear in $V$, the density in the center, 
$n(0) \approx \max(n_R,n_L)$ according to Eq.~(\ref{nfe}), obtains  for $L \gg |\ell_V|$ 
a correction proportional to $|V|$. As $j_c\approx eu n$, this implies a correction 
proportional to $|V| V$ to the current as soon as  $L \gg |\ell_V|$. Due to numerical 
problems, we were not able to obtain reliable numerical results in the small-$V$ 
regime $L \lesssim |\ell_V|$ where we expect a rounding of the non-analytic correction.  
Overall, the finite size corrections to the non-linear conductance are smaller than our 
numerical resolution  for $L \gg \ell_{\rm eq}$, see inset of Fig.~\ref{fig4}.

{\it Outlook.---} While our results have been derived only for 
spinless fermions with short ranged interactions, we expect that our 
main qualitative results are also of direct relevance for quantum 
wires made of electrons with spin and long-ranged Coulomb 
interactions. For these systems, the hydrodynamic equation 
(\ref{hydro}) should also be valid (with strongly modified parameters) 
 at least if gates provide screening. 
The highly asymmetric voltage drop predicted by us will probably be 
even much easier to observe, as the stronger interactions imply that 
the regimes $L \gg \ell_{\rm eq}$ and $L \gg |\ell_V|$ are much easier 
to reach. In recent experiments with ultracold atoms \cite{experiment} an atomic 
current was driven through a long quantum channel connecting two 
reservoirs and the resulting density profile (and therefore the drop 
of the chemical potential) was directly measured. This opens new
exciting possiblities to verify our predictions also in cold-atom 
experiments.

\acknowledgements We would like to thank E. Sela, M. Garst, G.~Zarand and  
especially J.~Rech and K. A. Matveev for useful discussions and acknowledge 
financial support by the DFG (SFB 608, FOR 960).

\clearpage

\title{Supplementary material to: ``Nonlinear conductance of long quantum wires 
at a conductance plateau transition: Where does the voltage drop?"}

\author{T.~Micklitz$^1$, A.~Levchenko$^2$, and A.~Rosch$^3$}

\affiliation{$^1$Dahlem Center for Complex Quantum Systems and
Institut f\"ur Theoretische Physik, Freie Universit\"at Berlin, 14195 Berlin, Germany \\
$^2$Department of Physics and Astronomy, Michigan State University, East Lansing, MI 48824, USA\\
$^3$Institut f\"ur Theoretische Physik, Universit\"at zu K\"oln, Z\"ulpicher Str. 77, 50937 Cologne, Germany}

\date{February 19, 2012}


\begin{abstract}

In this supplementary material we discuss (i) the irrelevance of short range 
interaction for spinless fermions, (ii) our numerical implementation 
of the Boltzmann equation, (iii) some more details on the calculation of the 
linear-response conductance and, finally, (iv) how a voltage drop along the 
quantum wire can be measured.

\end{abstract}

\maketitle

\subsection{Irrelevance of short ranged interactions at QCP}\vskip-.25cm

The imaginary-time action describing spinless fermion with short ranged interaction at the QCP ($\mu=0$) 
takes the form
\begin{align}
S=\int d\tau\int dx \left\{
\bar{\psi} \left(
\partial_\tau - {\hbar^2\over 2m} \partial^2_x
\right) \psi + V \bar{\psi} ( \partial_x \bar \psi )  ( \partial_x
\psi ) \psi
\right\}
\end{align}
The derivatives in the interaction term reflect Pauli's principle, 
$\psi^2=0$. From this action one can directly read off the scaling 
dimensions of the field, $[\psi]\sim L^{-1/2}$, and the (imaginary) 
time, $[\tau]\sim L^2$, for $[x]\sim L$. This simple power-counting 
shows that the interaction strength scales to smaller values, $V\to V/\lambda$, 
as the distance of Fermions is increased, $x\to \lambda x$, or as temperature is 
lowered, $T \to T/\lambda^2$. Therefore the interactions are irrelevant at the QCP. 
This result also implies that the quasiparticle picture and 
therefore the Boltzmann equation becomes asymptotically exact upon 
approaching the QCP.\vskip-.25cm

\subsection{Conservative splitting method}\vskip-.25cm
For the numerical implementation of the Boltzmann equation we follow 
Ref.~\cite{SAristov} and solve the time-dependent Boltzmann equation, by 
splitting time evolution into a free flow and a relaxation stage. 
The advantage of the splitting procedure is that the distribution 
obtained after a relaxation step can be corrected such that conservation 
laws are fulfilled in each collision, see Ref.~\cite{SAristov} for details.

Free flow and relaxation steps were implemented by a finite 
difference scheme, i.e. by discretizing two-dimensional phase-space. 
We used a first order implicit-explicit upwind scheme to model the 
free propagation step, and an implicit scheme for the relaxation 
step~\cite{SAristov}. The steady state is  reached when currents of 
the conserved quantities, $j_c$, $j_p$ and $j_E$ are constant along 
the wire. For the linear response calculation we parametrize 
$f_{x,p}=f^0_{p} +\delta f_{x,p}$, where $f^0_{p}$ is the 
distribution at zero bias and $\delta f_{x,p}$ is linear in $V$, and 
linearize the collision integral in $\delta f$. For the numerically 
more demanding calculations employing the full collision integral 
we used meshes with $22$  and $42$ points in momentum space and 
various different homogeneous and inhomogeneous discretization of 
space with $\sim100$ grid-points. We checked that our results are 
independent of discretization.

\subsection{Linear conductance in long wires}\vskip-.25cm
A method to calculate the linear conductance in long wires 
$\ell_{\rm eq}\ll L \ll |\ell_V|$, based on conservation laws and 
the distribution of fully equilibrated electrons Eq.~(6) in the main 
text, was originally introduced in Refs.~\cite{Sfeq,Speq}. As 
described in the main text, we need to calculate the ratios 
$r_1=j_E/j_c$, $r_2=\partial_x\delta T/\partial_x\delta\mu$ and 
$r_3=\partial_x j_E^R/\partial_x j_c^R$. The current $j_c$ in 
response to the applied voltage (to linear order $V$) is then found 
from
\begin{align}
\label{r3}
r_3={r_1j_c-j_E^0\over j_c-j_c^0}
\end{align} where $j_c^0$ and $j_E^0$  are the (linear response) 
charge and energy currents of non-interacting electrons which is 
directly described by Eq.~(2) of the main text,
 \begin{align}
j_c^0 &= {e^2V\over h} \alpha_0,
\quad
j^0_E = {eV\over h}
 \left(
\mu \alpha_0 + T \alpha_1
\right),
\quad
\alpha_k = \langle \xi^k \rangle_{z}
\end{align}
Here and in the following $\langle ... \rangle_{z}=  - \int_{-{z}}^{\infty} d\xi (...) {d f^0_\xi \over d\xi}$, 
$f^0_{\xi} ={1\over 1 + e^{\xi} }$, and $z= \mu/T$.

With these definitions we first calculate $r_1$. Using the 
equilibrium distribution Eq.~(6) of the main text 
we find
 \begin{align}
r_1={j_E\over j_c} = {1\over e} \left( \mu  + T\kappa \right),
\quad
\kappa =  { \langle \xi \sqrt{1 + \xi/ z} \rangle_{z}
\over
\langle \sqrt{1 +  \xi/ z }\rangle_{z}}
\end{align}

To calculate $r_2$ we use that momentum conservation implies 
homogeneity of momentum-current in the steady state. The latter can 
again be calculated with help of Eq.~(6) given in the main text by 
expanding $T(x)=T+\delta T(x)$, $\mu(x)=\mu+\delta \mu(x)$ form 
small $V$. To linear order in $V$, i.e., for small $\delta T$ and 
$\delta \mu$,  we find
\begin{align}
const. =
j_p(x) =  n \delta\mu(x)
+
n \kappa \delta T(x) + const.,
\end{align}
resulting in
\begin{align}
r_2={\partial_x \delta T\over \partial_x \delta\mu }= -\kappa^{-1}
\end{align}

To obtain $r_3$ we can directly use the definition of $j_c^R$ and 
$j_E^R$ given in the main text combined with the equilibrium 
distribution function~(6) of the main text
\begin{align}
\label{dnr}
\partial_x j^R_c
 =   {e \alpha_0\over h} \partial_x \delta\mu
+  {e \alpha_1\over h}  \partial_x \delta T
\end{align} and
\begin{align}
\label{der}
\partial_x j^R_E
=
 {\mu\over e} \partial_x j^R_c
+ {T\over h} \left( \alpha_1 \partial_x \delta\mu
+  \alpha_2 \partial_x \delta T \right).
\end{align} For the  ratio $r_2$ we therefore obtain
\begin{align}
r_3 = {\partial_x j^R_E\over \partial_x j^R_c}
= {\mu\over e} + {T\over e}{\alpha_1\kappa - \alpha_2\over \alpha_0\kappa - \alpha_1}
\end{align}
Inserting above expressions into \eqref{r3} and solving for $j_c$ 
gives Eq.~(13) of the main text.

\subsection{Measuring voltage profiles}\vskip-.25cm
To measure the voltage drop across a quantum wire one can, for 
example, use a weakly coupled tunneling contact realized by the tip of 
a scanning tunneling microscope. Assuming a constant 
tunneling matrix element $M$ and a constant density of states $\nu_0$ of the 
tunneling tip, the charge current $I_c(x)$ and the energy current 
$I_E(x)$ through the tip located at position $x$ are given by
\begin{eqnarray}\label{ic}
I_c(x)&=& {4\pi e\nu_0\over \hbar} |M|^2 \int dp\,  (f_{x,p} - f^{\rm tip}_p(\tilde \mu,\tilde T)) \\
I_E(x)&=& {4\pi e\nu_0\over \hbar} |M|^2 \int dp\,  \epsilon_p (
f_{x,p} - f^{\rm tip}_p(\tilde \mu,\tilde T) \nonumber
\end{eqnarray} 
Here $f_{x,p}$ is the distribution function of the wire and 
$f^{\rm  tip}_{x,p}=f^{\rm tip}_p(\tilde \mu(x),\tilde T(x))$ 
is the  Fermi distribution describing the 
occupation of the states in the tip. 
The latter is parametrized by the chemical potential $\tilde \mu$ and the temperature $\tilde T$.

The local chemical potential $\mu(x)$ of the quantum wire and the local 
temperature $T(x)$ of the wire are now obtained from the condition that 
particle and energy currents  flow only if there is a difference 
in the chemical potential and temperature. $T(x)$ and $\mu(x)$ 
are therefore obtained from the condition
\begin{eqnarray}\label{def}
T(x)=\tilde T, \quad \mu(x)=\tilde \mu \quad {\rm for}\ I_c(x)=I_E(x)=0
\end{eqnarray}
Note that this definition can 
be used for distribution functions far from equilibrium (and the usual 
result is obtained in equilibrium). 
Eq. (\ref{ic}) implies that the local chemical potential and 
temperature can directly be obtained from the local charge- and energy 
density of the system.

\begin{figure}[t!]
\begin{center}
\includegraphics[width=82mm]{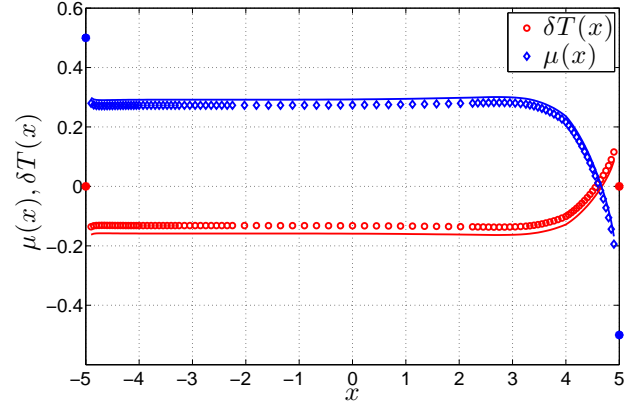}
\end{center}\vskip-.55cm
\caption{Local chemical potential $\mu(x)$ (blue) and temperature 
$T(x)=T+\delta T(x)$ (red) (in units $|eV|$) for a wire of length 
$L=10\ell_{\rm eq}$ and voltage $eV/T=0.4$. Symbols: Profiles 
obtained using Eq.~(\ref{ic},\ref{def}), i.e. for voltage contacts in the 
tunneling regime. Solid lines: Corresponding profiles obtained from an alternative fitting procedure (used in the main text) defined by Eq.~(\ref{defT2}). Both definitions give similar results for the parameters used in the paper.} \label{fig5}
\end{figure}

In Fig.~\ref{fig5} we show by the symbols chemical potential and temperature profiles obtained 
from the definition  (\ref{ic},\ref{def}) for a wire of length $L=10 \ell_{\rm  eq}$ 
with $eV/T=0.4$ at the QCP ($\mu=0$). The separate points 
at $\pm L/2$ denote the chemical potential in the two leads.

In the insets of Fig.~1 of the main text we also show $\mu(x)$ and 
$T(x)$ but in this case we use a {\em different} definition of these 
quantities, which is more appropriate to illustrate our analytical 
arguments. 
In the main text, we fit the local charge, energy {\em and} momentum 
densities to the equilibrium distribution given by Eq.~(6) of  the main text, which depends not only $\mu$ and $T$ but also on 
the average velocity $u$. At each point $x$, we therefore define the three space-dependent functions $u(x),\mu(x)$, and $T(x)$ by 
\begin{eqnarray}\label{defT2}
\sum_p  f_{x,p}&=&\sum_p  f^{\rm eq}_p(u,\mu,T) \\
\sum_p \epsilon_p  f_{x,p}&=&\sum_p   \epsilon_p  f^{\rm eq}_p(u,\mu,T) \nonumber \\
\sum_p p  f_{x,p}&=&\sum_p p  f^{\rm eq}_p(u,\mu,T) \nonumber
\end{eqnarray}
The two definitions, (\ref{ic},\ref{def}) and
(\ref{defT2}), are different, as for the tunneling tip momentum is not conserved and we fit in the first case to an equilibrium function $f^{\rm tip}$ which depends only on the chemical potential and the temperature but not on the velocity.

For the range of applied voltages discussed in this paper, however, both 
methods lead to nearly identical profiles. This is shown in 
Fig.~\ref{fig5}, where temperature and chemical-potential profiles 
obtained from Eqs.~ (\ref{ic},\ref{def}) (symbols) are compared to the 
corresponding curves from Eq.~(\ref{defT2}) (lines) which are also used in the main text. 

In conclusion, we have discussed an experimental procedure which allows to measure the local chemical potential and local temperature by using tunneling contacts. While we used in the main text a different definition (the one needed for our analytic arguments), the resulting profiles are almost identical which guarantees that the voltage and temperature profiles shown in the main text can be measured.


\begin{thebibliography}{99}
\bibitem{fnW}
For a short-ranged interaction potential $V(x)=V_0e^{-q_0|x|/\hbar}$ we find $W={2\pi\over\hbar} \left({9m \hbar^2 \over 2\pi^2L^2}{V_0^2\over q_0^{10}}\right)^2$.
\end{thebibliography}
\end{document}